# Complete coherent control of spin qubits in self-assembled InAs quantum dots under oblique magnetic fields

I. Samaras,[1, *] K. Barr,[1] C. Schneider,[2] S. Höfling,[3] and K.G. Lagoudakis[1, †]

[1]*University of Strathclyde, Glasgow G4 0NG, United Kingdom*
[2]*Institut für Physik, Fakultät V, Carl von Ossietzky Universität Oldenburg, 26129 OLdenburg, Germany*
[3]*Julius-Maximilians-Universität Würzburg, Technische Physik, 97074 Würzburg, Germany*

We demonstrate complete coherent control of a single spin qubit confined in a self-assembled InAs negatively charged quantum dot subjected to an Oblique magnetic field, and directly compare this regime with the conventional Voigt geometry. In the Oblique-field configuration, the ground-state spin eigenstates are found to be unequal superpositions of the bare electron spin, with their composition tunable via the orientation of the applied field. This tunable spin mixing provides an additional degree of freedom to engineer the spin basis and associated optical couplings in the charged quantum dot system. Although this geometry has a distinct structure with important implications, it provides a regime in which we can fully and coherently control the tailored spin qubit. We observe Rabi oscillations and Ramsey fringes, and demonstrate arbitrary single-qubit rotations, enabling a direct comparison with the Voigt case. Our results establish that spin-qubit control does not necessarily require a pure Voigt geometry and can instead be achieved under Oblique magnetic fields. This relaxes constraints on device and field alignment and offers a versatile route to design and optimize spin–photon interfaces and spin-based quantum information processing architectures in semiconductor quantum dots.

## I. INTRODUCTION

Self-assembled quantum dots (QDs) are a leading solid-state platform for optical quantum technologies. Strong three-dimensional confinement produces discrete, atom-like energy levels with large oscillator strengths and near transform-limited optical transitions, enabling an extremely broad range of applications. When charged, these QDs host a single electron or hole spin that forms a long-lived ground-state qubit, optically accessed via trion transitions with well-defined selection rules for initialization, control, and readout [1]. In addition, self-assembled QDs have supported a broad range of single-spin protocols, including spin pumping [2], coherent control [3–15], spin echo techniques [16], spin-photon interface [17, 18], spin-photon entanglement [19, 20] and are typically demonstrated in standard magnetic-field orientations like the Faraday [21] or Voigt configuration. In these standard geometries, coherent control is well established, and performance is typically constrained by acoustic phonon interactions [22], and charge noise [23] rather than the control mechanism itself. Magnetic fields tilted away from the Faraday or Voigt configurations provide additional degrees of freedom by exploiting traits from both geometries. This additional tunability is attractive because Faraday fields provide near-cycling transitions suited to high-fidelity optical readout, whereas Voigt fields yield the double-Λ structure commonly used for optical spin control. Related progress has recently been reported in Faraday geometry, where light-hole mixing is used to realize a highly asymmetric Λ-system with cyclic optical transitions for spin control and readout [12], whereas in this work we instead use an oblique magnetic field to engineer the spin basis and demonstrate universal all-optical control in a tilted-field regime. Thus, an Oblique field, in principle, allows one to interpolate between these regimes and balance "enough cycling" for readout with "enough mixing" for control in a single static configuration [24]. Despite this promise, demonstrations of complete coherent control of single QD spin qubits, such as arbitrary SU(2) rotations validated by Rabi and Ramsey measurements, have largely been performed in the Voigt configuration, where the selection rules simplify giving rise to a double-Λ system with orthogonal linear polarizations for each branch [24, 25]. By contrast, tilted-field studies have focused predominantly on extracting g-tensor anisotropies [26–28], or on manipulating excitonic (bright/dark) manifolds [29–31], rather than investigating the resulting tailored ground-spin states [24] as an optically controllable qubit.

Here we close this gap by demonstrating complete coherent control of a charged self-assembled InAs QD in a fixed Oblique magnetic field and compare it with the Voigt data acquired from the same QD. We show that a single pulsed laser can implement arbitrary qubit rotations even when the field is tilted away from both the Faraday and Voigt orientations. Despite the fact that the ultrafast rotations happen around an axis that is tilted away from the equatorial plane of the Bloch sphere, we observe clear Rabi oscillations and by carefully tuning the pulse power of a dual pulse train to the correct rotation angle, we observe Ramsey fringes. Finally, combining two rotations separated by a precisely timed Larmor precession, we realize general SU(2) rotations on the Bloch sphere. Our experimental findings are complemented by quantum optical simulations that capture the full com-plexity of the Oblique field and show excellent agreement with the measured data. These results establish that universal all-optical single-qubit control is compatible with Oblique magnetic fields in self-assembled QDs, linking the mature Voigt-geometry toolbox to the tilted-field configuration with its additional degrees of freedom.

---

* ioannis.samaras@strath.ac.uk
† k.lagoudakis@strath.ac.uk

## II. EXPERIMENTAL DETAILS

### A. Sample, cryogenic environment & magnet geometry

The experiments are performed in a self-assembled $\delta$-doped InAs quantum dot (QD) sample grown by molecular beam epitaxy in Stranski–Krastanow mode and positioned at the center of a GaAs microcavity with 5 (18) top (bottom) Distributed Bragg Reflector AlAs/GaAs mirror pairs. A single InAs QD layer is capped by 100 nm of GaAs, a sample which is discussed in detail in Ref. [32]. Charging is promoted by a Si $\delta$-doped layer (areal density $\sim 10^{11}\,\mathrm{cm}^{-2}$) placed 10 nm below the QD layer, so that approximately one third of the dots are singly charged. The QDs investigated emit in the $910-920\,\mathrm{nm}$ range. The specific dot studied here is selected under above-band (Ab. Bd.) excitation at 780 nm by comparing spectra with and without the applied magnetic field and choosing a dot that exhibits the characteristic fourfold Zeeman splitting, an indication for charged QDs. The sample is mounted in a continuous-flow liquid-helium magneto-cryostat (Oxford Microstat-MO) with a helium cryocirculator (ColdEdge Stinger), enabling stable closed-cycle operation while allowing the use of the integrated superconducting solenoid to generate a strong magnetic field [33]. The system provides a homogeneous magnetic field up to 5 T along the cryostat axis. Unless otherwise stated, measurements are performed at $T = 5.0 - 5.5\,\mathrm{K}$. For Zeeman-splitting measurements, the temperature is set to 6.5 K to reduce spectral drift during magnetic-field ramps caused by magnet/sample heating. To access the two magnetic field configurations studied here, two different custom-built sample holders are used. For the Oblique geometry, the sample holder orients the sample at $\vartheta = 60°$ from Faraday geometry, while optical access is enabled using an appropriately tilted mirror in close proximity to the sample. The same principle is used for the Voigt geometry, where the angle is at $\vartheta = 90°$ as shown in Fig. 1(a) [24]. The angle $\vartheta$ is defined with respect to the growth axis (Faraday).

### B. All optical excitation & detection scheme

Optical excitation and collection are performed with a $0.5 - \mathrm{NA}$ long-working-distance microscope objective (W.D. = 12 mm) producing a $\sim 1\,\mu\mathrm{m}$ (FWHM) spot on the sample. As the sample is static, we can access different areas of the sample surface by moving the mi-croscope objective. We use a weak above band (Ab. Bd.) laser at 780 nm for initial characterization, which is switched off during the main coherent control experiment. For all-optical coherent control we use pulses from an ultrafast Ti:sapphire laser with a repetition rate of at 80.2 MHz. The pulse duration in the sample is 3 ps (intensity FWHM), short compared to the trion lifetime ($\sim 0.8\,\mathrm{ns}$). We spectrally shape and filter the off-resonant pulses so that their spectral centers are red-detuned by approximately $\sim 500\,\mathrm{GHz}$ from the targeted $\Lambda$-system under investigation, thereby minimizing population transfer to the trion state [34]. For spin initialization and readout we use a weak resonant CW laser which resonantly drives the highest energy outer transition $|\uparrow\rangle_{obl} \to |\uparrow\downarrow\Uparrow\rangle_{obl}$ of the QD, whereas in Voigt geometry the corresponding highest energy outer transition of the Voigt-basis $\Lambda$-system is $|\uparrow\rangle_v \to |\uparrow\downarrow\Uparrow\rangle_v$. The incident polarization and power of the resonant CW laser is set using a combination of a motorized polarizer, half- and quarter-waveplates. To minimize the effect of the CW resonant laser during the coherent control experiment while maintaining high enough detected photon counts, we set its power in the range of $50 - 150\,\mathrm{nW}$ with the latter being 18 times lower than the value where we can distinguish any observable dressing of the transition (see Suppl. Info.). Residual back-reflected laser light from the resonant CW laser, is rejected by spectral filtering in the spectrometer. For readout we isolate spontaneously emitted photons from the $|\uparrow\downarrow\Uparrow\rangle_{obl} \to |\downarrow\rangle_{obl}$ ($|\uparrow\downarrow\Uparrow\rangle_v \to |\downarrow\rangle_v$) transition in the Oblique (Voigt) configuration by spectral filtering through a double 1 m long custom-built spectrometer with intermediate slits that has $\sim 50\%$ efficiency at maximum resolution ($\sim 8.5\,\mu\mathrm{eV}$), while photon detection is performed using a single photon counting module (SPCM) with $\sim 150\,\mathrm{s}^{-1}$ dark counts at the output of the spectrometer. The location of the pulses with respect to the PL spectrum of the charged quantum dot in the Oblique field configuration is shown in Fig. 1(b), where the shaded areas on the highest energy transitions highlight the resonantly driven (grey) and detected (yel-low) transitions.

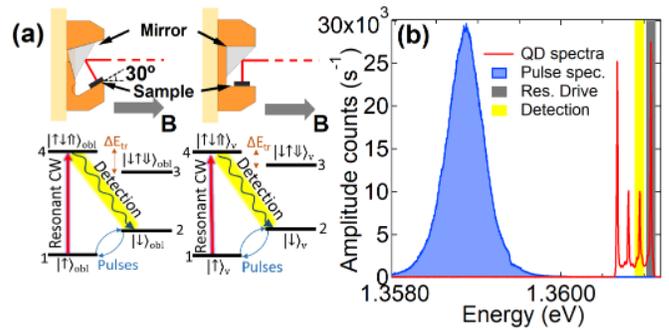

FIG. 1. (a) Sample holders for Oblique (left) and Voigt (right) along with the double-$\Lambda$ systems for each one of the geometries. (b)Raw PL shows the excitation/detection scheme of our QD with the non-resonant pulses in the Oblique geometry.

## C. Zeeman Splitting interaction & influence on the selection rules

We address a singly charged InAs QD hosting a resident electron spin in the ground state and a trion in the excited state. The magnetic field dependence of the optical transition energies between the trion and ground-spin state manifolds reflects the interplay of the linear in field Zeeman interaction and a quadratic diamagnetic shift. We investigate two magnetic field configurations, with the applied magnetic field lying in the x-z plane, at angles $\vartheta_1 = 60°$, and $\vartheta_2 = 90°$ with respect to the growth axis z, corresponding to the Oblique and Voigt geometries, respectively. Electron and heavy-hole Zeeman interactions are described by effective spin-1/2 Hamiltonians with different in-plane and out-of-plane g-factors. The Oblique magnetic field mixes the z-basis spins and generates unequal superpositions for the ground spin- and excited trion- states which we label as $|\uparrow\rangle_{obl}$, $|\downarrow\rangle_{obl}$, and $|\uparrow\downarrow\Uparrow\rangle_{obl}$, $|\uparrow\downarrow\Downarrow\rangle_{obl}$. The composition of these states in terms of the z-basis spins is directly related to the magnetic field angle and the in-plane ($g^V$) and out-of-plane ($g^F$) g-factors [24]. From the fitted magnetophotoluminescence spectra in Fig. 2(a) taken in the Oblique configuration, we extract the effective g-factors $g_e^{eff} = 0.459$ and $g_h^{eff} = 0.919$. Combining this with the magnetophotoluminescence spectra in the Voigt configuration in Fig. 2(b), we extract both the in-plane and out-of-plane g-factors for the electron and the heavy-hole. The electron is found to be only weakly anisotropic since $g_e^F = 0.497$ and $g_e^V = 0.446$, whereas the hole g-factor has a more pronounced anisotropy with $g_h^F = 1.823$ and $g_h^V = 0.129$. This anisotropy is particularly important in the Oblique-field geometry, where the in-plane component of the magnetic field induces heavy-hole mixing, thereby renormalizing the hole response and modifying the optical selection rules. Because the electron g-factor is only weakly anisotropic, the effective Zeeman field remains nearly collinear with the applied magnetic field. By contrast, the stronger anisotropy of the heavy-hole primarily affects the trion eigenstates, setting their Zeeman splitting and, crucially, determining the optical coupling strengths and transition polarizations of the Oblique-field double-Λ system. The four allowed optical transitions connect the two ground electron eigenstates to the two excited trion eigenstates and inherit their polarization properties from the optical selection rules in the applied magnetic field (Fig. 2(c)). In the Voigt configuration, the corresponding polarization properties are shown in Fig. 2(d). More generally, the Oblique-field selection rules differ from the purely circular Faraday limit and the purely linear Voigt limit, and are sensitive to the underlying spin mixing [24]. At 5T spectrally- and polarization-resolved photoluminescence is used to determine the optical coupling strengths and transition polarizations of the Oblique-field double Λ-system. We measure the polarization properties using rotating-quarter-waveplate polarimetry finding that all four transitions are circularly polarized with high ($\geq 95\%$) circularity (Fig. 2(c)). Finally, we report that after following the same procedure to extract the weights in the Voigt geometry, all four transitions are found to be linearly polarized (Fig. 2(d)).

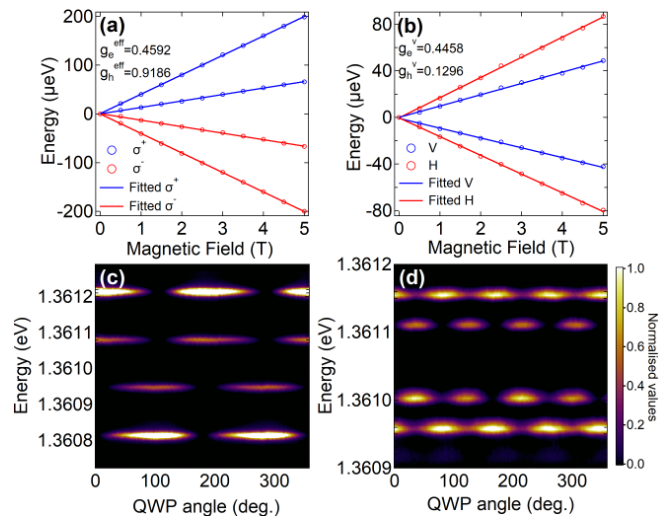

FIG. 2. Zeeman effect (a) in the Oblique configuration and (b) in the Voigt configuration. polarization analysis for each emission peak at 5T (a) In Oblique field geometry the two higher energy transitions are $\sigma^+$ polarized whereas the two lower energy transition $\sigma^-$ (b) In Voigt geometry, the outer transitions become horizontally polarized while the inner transitions vertically polarized. Figures (c), (d) share the same colorbar.

## III. RESULTS

### A. Rabi Oscillations

Coherent single-qubit rotations are evidenced by Rabi oscillations under non-resonant pulsed driving, an approach widely used in QD spin-qubit control experiments [3–15]. In both magnetic field orientations, the qubit is encoded in the ground state electron spin of a negatively charged QD, while far red-detuned picosecond (ps) pulses couple these ground states to the trion manifold only virtually [34]. Because the trion is negligibly populated during the pulse, the interaction can be described as an effective AC-Stark coupling, which induces a coherent mixing of the spin states together with an additional AC-Stark phase [34]. As a result, the pulse implements a spin rotation by an angle that scales approximately as the square root of the pulse power, $\theta \propto \sqrt{P}$. The key difference between the two magnetic field configurations investigated here, lies in the orientation of the axis about which the spin qubit rotates. In the Oblique-field configuration, we label the ground states, as $|\uparrow\rangle_{obl}$ and $|\downarrow\rangle_{obl}$. On the Bloch sphere, a single red-detuned circularly po-larized picoseconds pulse, cross-polarized with respect to the relevant $\Lambda$-system, drives rotations about a tilted axis $\hat{n}$, whose orientation is determined by the angle of the applied magnetic field, assuming weak electron g-factor anisotropy (Inset Fig. 3(a)). By contrast, in Voigt geometry, the ground states are $|\uparrow\rangle_v$ and $|\downarrow\rangle_v$, so the Bloch vector rotates about the $\hat{x}$ axis defined by the magnetic field, which is perpendicular to the growth axis $\hat{z}$ (Inset Fig. 3(b)). For both magnetic field orientations we therefore see that for increasing pulse powers the detected signal shows a clear oscillatory behavior with damping, indicating effective Rabi rotations of the qubit about the respective rotations axes.

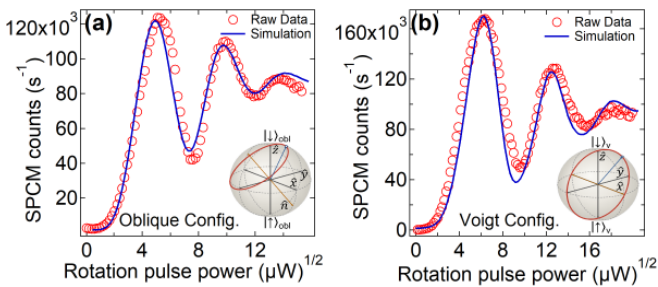

FIG. 3. (a) Rabi oscillations in Oblique configuration as a function of pulse power. Experimental data are shown as red markers and the corresponding simulation is shown as a blue solid line. Inset: Bloch Sphere representation of the driven Bloch-vector as it is rotated about the axis $\hat{n}$, tilted by 60° deg. (b) Rabi oscillations in Voigt configuration. Inset: Bloch Sphere representation at a field angle of 90°, where the Bloch vector is rotated about the $\hat{x}$ axis.

The resulting signal for the Oblique (Voigt) configuration, collected from the $|\uparrow\downarrow\Uparrow\rangle_{obl} \to |\downarrow\rangle_{obl}$ ($|\uparrow\downarrow\Uparrow\rangle_v \to |\downarrow\rangle_v$) transitions, is plotted in Fig. 3(a) (3(b)) as a function of $\sqrt{P}$ of the non-resonant pulses. The damping observed in both geometries is consistent with a combination of excitation-induced dephasing, phonon-assisted relaxation, spontaneous emission, and Larmor precession [22]. To quantify these effects, we compare the measurements with master-equation simulations of the driven four-level system implemented in QuTiP [35, 36]. The model includes spontaneous emission, pure dephasing, and excitation-induced dephasing arising from coupling to longitudinal acoustic phonons, which produces pulse-area-dependent Rabi oscillation damping in InAs/GaAs quantum dots [22]. The parameters extracted from these fits are then used consistently in simulations of red-detuned pulse-driven Rabi oscillations, Ramsey interference, and general SU(2) control.

### B. Ramsey Interference

Following the demonstration of Rabi oscillations, we probe the ground-state coherence using a Ramsey sequence. In both geometries, the sequence consists of two identical control pulses separated by a variable delay $\tau$, with each pulse chosen to implement an effective $\pi/2$ rotation that brings the Bloch vector to the equator. The key distinction is that, in the Oblique-field configuration, where the magnetic field is oriented at $\vartheta = 60°$ with respect to the growth axis, this effective $\pi/2$ operation requires an actual pulse rotation angle of $\theta \approx 109°$. By contrast, in the Voigt geometry, corresponding to 90°, the effective $\pi/2$ condition coincides with the physical $\pi/2$ pulse. The Rabi oscillation data allow us to identify the pulse powers corresponding to these operating points. We therefore use two identical $\theta = 109°$ pulses in the Oblique-field Ramsey experiment and two identical $\theta = 90°$ pulses in the Voigt case. In the Oblique-field configuration, the first pulse rotates the Bloch vector from the north pole to the equatorial plane about the fixed control axis $\hat{n}$, thus preparing a coherent superposition of $|\uparrow\rangle_{obl}$ and $|\downarrow\rangle_{obl}$ (inset of Fig. 4(a)). During the free-evolution interval, the spin precesses about the $\hat{z}$ axis, at the Larmor frequency $\omega^L_{obl}/2\pi = 32\,\mathrm{GHz}$, leading to an overall rotation by an angle $\phi = \omega\tau$ [5–11, 15]. The second pulse then performs a further rotation about the same control axis $\hat{n}$, mapping the position on the equator into a delay-dependent ground-state superposition. As in the Rabi-oscillation measurements, the final population is read out through spontaneous spectrally filtered emission on the $|\uparrow\downarrow\Uparrow\rangle_{obl} \to |\downarrow\rangle_{obl}$ transition, yielding Ramsey fringes as a function of $\tau$ (Fig. 4(a)). In this representation, the maximum signal corresponds to a superposition near the $|\uparrow\rangle_{obl}$ state, while the minimum signal corresponds to a superposition near the $|\downarrow\rangle_{obl}$ as illustrated in the inset of Fig. 4(a). The corresponding Ramsey measurement in Voigt geometry follows the same general protocol, but with control pulses implementing $\theta = 90°$ rotations, creating coherent superpositions between $|\uparrow\rangle_v \leftrightarrow |\downarrow\rangle_v$, consistent with the different orientation of the rotation axis in that configuration (Fig. 4(b)). The Larmor frequency here is $\omega^L_v/2\pi = 31\,\mathrm{GHz}$, consistent with the slightly smaller electron ground state splitting. The measured fringes are well reproduced by our master-equation model using the parameters extracted from the Rabi-oscillation analysis. The delay $\tau$ is controlled with a Mach–Zehnder interferometer [37] in the pulsed-excitation path, where a motorized translation stage provides coarse delay steps of 1.33 ps, spanning delays from $\tau = 0$ up to 1 ns.

An interesting difference between the two configurations is the initial phase offset $\Delta\phi$ of the Ramsey fringes (here $\Delta\phi \cong 0.24\,\mathrm{rad}$). Neglecting Larmor precession during the all-optical rotations, in the Voigt configuration, two successive $\pi/2$ pulses bring the Bloch vector from the north pole to the south pole of the Bloch sphere, yielding maximum counts and therefore a zero Ramsey-fringe phase. In the oblique configuration, the effective $\pi/2$ pulses rotate the Bloch vector about the tilted control axis $\hat{n}$. When applied with zero interpulse delay, these pulses drive a Rabi-like trajectory about $\hat{n}$ (Inset of Fig. 3(a)). Consequently, the south pole is reached only for a small delay between the pulses, which introduces a finite phase shift in the Ramsey fringes. This phase offset is therefore a direct consequence of the tilted control axis $\hat{n}$.

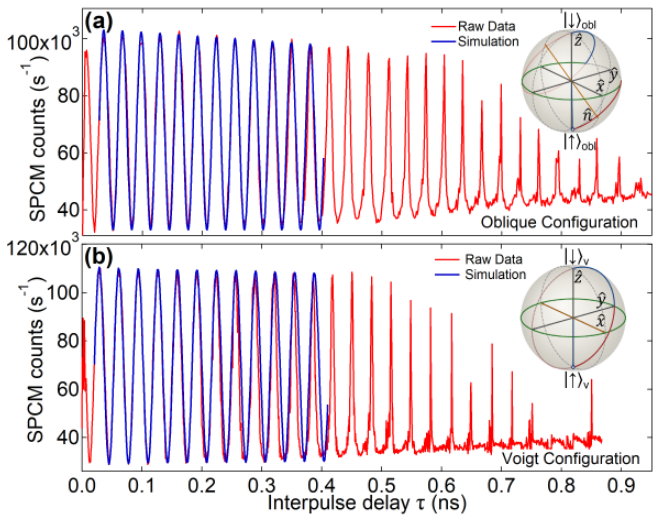

FIG. 4. Ramsey interference measured by driving the qubit with two effective $\pi/2$ pulses separated by an interpulse delay in (a) Oblique and (b) Voigt configurations. The delay here is stepped in 1.3 ps increments, producing delay dependent oscillations. The red solid line denotes the experimental data and the solid blue line is our modelled Ramsey interference. At zero delay the noise is because of optical interference between the pulses. At larger delays, signatures consistent with nuclear-spin polarization become clearly visible. The insets show a simplified state trajectory for two pulses with optical area of two $\pi/2$ in the Voigt and pulses corresponding to $\theta = 109°$ in the Oblique configuration, at a delay, for which the second pulse brings the Bloch vector to the opposite spin state.

At longer delays, deviations from simple oscillatory behavior appear in the form of pronounced spikes in the signal, consistent with signatures of dynamic nuclear-spin polarization reported previously [38–41]. Interestingly, the onset of this behavior seems to happen for shorter delays in the Voigt configuration. Strategies to mitigate nuclear-spin and charge-noise limitations, and thereby extend spin coherence in self-assembled quantum dots, have been extensively explored elsewhere [42].

### C. Complete coherent control

As a final step, we demonstrate complete coherent control in an Oblique magnetic field, namely universal single-qubit operations on the ground $|\uparrow\rangle_{obl} \leftrightarrow |\downarrow\rangle_{obl}$ manifold. The protocol generalizes the Ramsey sequence by using two identical control pulses with tunable pulse areas and delays. In this setting, the first pulse applies a coherent rotation about the control axis $\hat{n}$, while Larmor precession during the interpulse delay $\tau$ rotates the qubit by an angle $\phi = \omega\tau$ about the $\hat{z}$ axis. The second pulse then induces a further rotation about the axis $\hat{n}$. Two rotations about the fixed control axis $\hat{n}$, separated by an additional controlled rotation about the $\hat{z}$ axis, are sufficient to reach any point on the Bloch sphere, and therefore implement arbitrary SU(2) single-qubit rotations. The experiment is performed at a non-zero pulse delay ($\tau \approx 20 - 140$ ps) to avoid optical-field interference between the two pulses. As in the previous experiments, for each pulse power, which sets the rotation angle, we scan the delay $\tau$ and record time-integrated photon counts from the $|\uparrow\downarrow\Uparrow\rangle_{obl} \to |\downarrow\rangle_{obl}$ transition. The resulting data form a two-dimensional control map with axes given by $\sqrt{P}$ and delay $\tau$. The measured map exhibits the characteristic multi-lobed structure expected for coherent SU(2) control in a detuned $\Lambda$-systems [5, 8, 9, 15] (Fig. 5(a)). However, in the present case, the multi-lobe pattern also displays distinctive features, including tilted and "interconnected" lobes. A master-equation model, using parameters independently constrained by the Rabi and Ramsey measurements, reproduces well the observed lobe pattern (Fig. 5(b)). Finally, we follow a similar procedure to reproduce the complete coherent control of the $|\uparrow\rangle_v \leftrightarrow |\downarrow\rangle_v$ manifold in the Voigt configuration, as indicated in Fig. 5(c) and its modelled results in Fig. 5(d). Unlike the Voigt configuration, the Oblique-field configuration yields a tilted control axis and asymmetric couplings to the two ground states. Therefore, one would expect that these differences should visibly modify the SU(2) control maps between the two configurations. Interestingly, we show that the SU(2) control maps show minor differences mostly related to the different Larmor frequencies and the slightly modified pulse powers due to asymmetric couplings. Overall, these results establish that self-assembled InAs quantum dots in a fixed Oblique magnetic field support universal all-optical single-qubit control, while preserving the tunability of the spin eigenstates through the field orientation.

## IV. SIMULATIONS

We model the QD as a four-level double-$\Lambda$ system comprising two ground spin states, $|1\rangle$ and $|2\rangle$, split by the ground state splitting $\Delta E_{gs}$, and two excited states, $|3\rangle$ and $|4\rangle$, separated by the excited state splitting $\Delta E_{es}$. The dynamics is governed by a Lindblad master equation:

$$\dot{\rho} = -i[H, \rho] + \sum_m \mathcal{L}(c_m), \quad (1)$$

where $\mathcal{L}(c_m)$, the Lindblad superoperators of the collapse operators $c_m$. All calculations are performed in a frame rotating at the continuous-wave (CW) laser frequency $\omega_l$, within the rotating-wave approximation. The unperturbed Hamiltonian in this frame is

$$H_0 = \frac{\Delta E_{gs}}{2}(p_{22} - p_{11}) + \Delta_{cw} p_{33} + (\Delta_{cw} + \Delta E_{es}) p_{44}, \quad (2)$$

where $p_{ff} = |f\rangle\langle f|$ is the population projection operator onto the states, and $\Delta_{cw}$ is chosen to render the CW

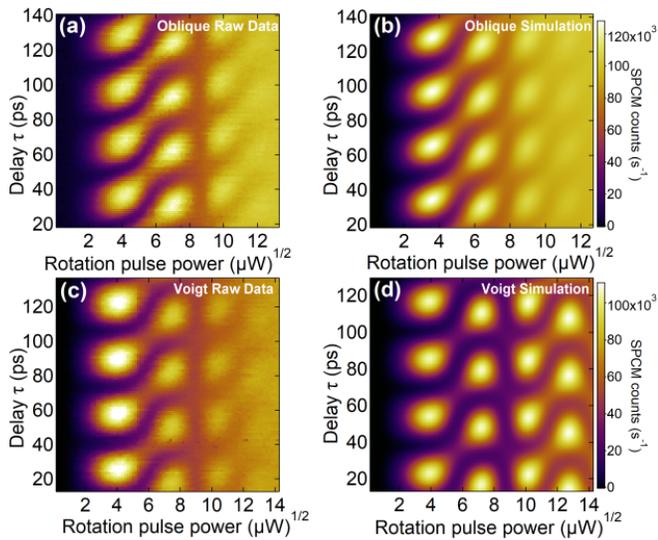

FIG. 5. Detected photon counts following excitation with two off-resonant pulses of variable power and variable interpulse delay. Scanning these parameters enables access to the full Bloch Sphere. (a) Experimental data for the Oblique case, (b) Simulated data in Oblique. (c) Experimental data for the Voigt case. (d) Simulated data for the Voigt. Figures (a) and (b) as well as (c) and (d) share the same colorbar.

laser resonant with the outer $|1\rangle \leftrightarrow |4\rangle$ transition. The Hamiltonian of a CW field of Rabi frequency $\Omega_{\text{cw}}$ that drives this transition time-independently in the rotating frame is:

$$H_{\text{cw}} = \tfrac{\Omega_{\text{cw}}}{2} k_{14} \left(p_{41} + p_{41}^{\dagger}\right), \qquad (3)$$

where $k_{14}$ is the coupling dipole strength in each configuration that we extract from (Fig. 2(c),(d)). A pulsed field with Gaussian envelope $\Omega(t) = \Omega_p \exp[-(t-t_0)^2/2\sigma^2]$ is decomposed into coefficients that are responsible for the driving of each transition, based on the polarization in each geometry:

$$\left.\begin{array}{l}\Omega_p^{14}(t) = \Omega_p^{24}(t) = \Omega(t)\cos\alpha, \\ \Omega_p^{13}(t) = \Omega_p^{23}(t) = \Omega(t)e^{i\beta}\sin\alpha\end{array}\right\} \text{Oblique}$$
$$\left.\begin{array}{l}\Omega_p^{14}(t) = \Omega_p^{23}(t) = \Omega(t)\cos\alpha, \\ \Omega_p^{13}(t) = \Omega_p^{24}(t) = \Omega(t)e^{i\beta}\sin\alpha\end{array}\right\} \text{Voigt}$$
(4)

where $\alpha$ and $\beta$ are set to correspond to the polarization and relative phase, so that the pulses drive the transitions accordingly for each configuration. The pulse carrier is detuned from the CW frame by $\Delta_p$. The full pulse Hamiltonian is

$$H_p(t) = \frac{1}{2}\sum_{i,j} k_{ij} \Omega_p^{ij}(t) e^{-i\Delta_p t} p_{ij} + \text{H.C} \qquad (5)$$

where, $i(j)$ are the ground (excited) state indices and $k_{ij}, \Omega_p^{ij}$ the coupling dipole strengths and pulse coefficients for every transition in each configuration. Radiative decay from each excited state to each ground state is included through collapse operators $c_{ji} = \sqrt{\Gamma_{ji}}\, p_{ji}$ with $\Gamma_{ji}^{-1}$ the trion lifetime. Pure dephasing is modeled by static Lindblad operators $\sqrt{\gamma_d}\, p_{jj}$, representing charge noise, supplemented by a time-dependent phonon term $\sqrt{\alpha_{phonon}}\,|\Omega_p(t)|p_{jj}$. Starting from $\rho(0) = |2\rangle\langle 2|$, we compute the time-integrated photon flux on the $|4\rangle \to |2\rangle$ channel,

$$N = \int \Gamma_{42} \langle p_{44} \rangle \, \mathrm{d}t, \qquad (6)$$

as a function of pulse area, using the QuTiP master equation solver. The simulation parameters can be found in Ref. [43].

## ACKNOWLEDGMENTS

I.S. acknowledges financial support from the EPSRC Doctoral Training Partnership under grant no. EP/W524670/1. K.B. acknowledges financial support from the EPSRC Doctoral Training Program under grant no. EP/R513349/1. C.S. is grateful for financial support by the BMFTR within the project COMPHORT (16KIS2107). Comphort is funded within the QuantERA programm.

[1] A. J. Ramsay, S. J. Boyle, R. S. Kolodka, J. B. B. Oliveira, J. Skiba-Szymanska, H. Y. Liu, M. Hopkinson, A. M. Fox, and M. S. Skolnick, Fast optical preparation, control, and readout of a single quantum dot spin, Phys. Rev. Lett. **100**, 197401 (2008).

[2] B. D. Gerardot, D. Brunner, P. A. Dalgarno, P. Öhberg, S. Seidl, M. Kroner, K. Karrai, N. G. Stoltz, P. M. Petroff, and R. J. Warburton, Optical pumping of a single hole spin in a quantum dot, Nat. **451**, 441 (2008).

[3] M. H. Mikkelsen, J. Berezovsky, N. G. Stoltz, L. A. Coldren, and D. D. Awschalom, Optically detected coherent spin dynamics of a single electron in a quantum dot, Nat. Phys. **3**, 770 (2007).

[4] J. Berezovsky, M. H. Mikkelsen, N. G. Stoltz, L. A. Coldren, and D. D. Awschalom, Picosecond coherent optical manipulation of a single electron spin in a quantum dot,

Sci. **320**, 349 (2008).

[5] D. Press, T. D. Ladd, B. Zhang, and Y. Yamamoto, Complete quantum control of a single quantum dot spin using ultrafast optical pulses, Nat. **456**, 218 (2008).

[6] E. D. Kim, K. Truex, X. Xu, B. Sun, D. G. Steel, A. S. Bracker, D. Gammon, and L. J. Sham, Fast spin rotations by optically controlled geometric phases in a charge-tunable InAs quantum dot, Phys. Rev. Lett. **104**, 167401 (2010).

[7] J. A. Gupta, R. Knobel, N. Samarth, and D. D. Awschalom, Ultrafast manipulation of electron spin coherence, Sci. **292**, 2458 (2001).

[8] K. De Greve, P. L. McMahon, D. Press, T. D. Ladd, D. Bisping, C. Schneider, M. Kamp, L. Worschech, S. Höfling, A. Forchel, and Y. Yamamoto, Ultrafast coherent control and suppressed nuclear feedback of a single quantum dot hole qubit, Nat. Phys. **7**, 872 (2011).

[9] K. D. Greve, D. Press, P. L. McMahon, and Y. Yamamoto, Ultrafast optical control of individual quantum dot spin qubits, Rep. Prog. Phys. **76**, 092501 (2013).

[10] T. M. Godden, J. H. Quilter, A. J. Ramsay, Y. Wu, P. Brereton, S. J. Boyle, I. J. Luxmoore, J. Puebla-Nunez, A. M. Fox, and M. S. Skolnick, Coherent optical control of the spin of a single hole in an InAs / GaAs quantum dot, Phys. Rev. Lett. **108**, 017402 (2012).

[11] M. R. Hogg, N. O. Antoniadis, M. A. Marczak, G. N. Nguyen, T. L. Baltisberger, A. Javadi, R. Schott, S. R. Valentin, A. D. Wieck, A. Ludwig, and R. J. Warburton, Fast optical control of a coherent hole spin in a microcavity, Nat. Phys. **21**, 1475 (2025).

[12] Z. X. Koong, U. Haeusler, J. M. Kaspari, C. Schimpf, B. Dejen, A. M. Hassanen, D. Graham, A. J. Garcia, M. Peter, E. Clarke, M. Hugues, A. Rastelli, D. E. Reiter, M. Atatüre, and D. A. Gangloff, Coherent control of quantum-dot spins with cyclic optical transitions, arXiv preprint arXiv:2509.14445 (2025).

[13] W. B. Gao, A. Imamoglu, H. Bernien, and R. Hanson, Coherent manipulation, measurement and entanglement of individual solid-state spins using optical fields, Nat. Photon. **9**, 363 (2015).

[14] D. Ding, M. H. Appel, A. Javadi, X. Zhou, M. C. Löbl, I. Söllner, R. Schott, C. Papon, T. Pregnolato, L. Midolo, A. D. Wieck, A. Ludwig, R. J. Warburton, T. Schröder, and P. Lodahl, Coherent optical control of a quantum-dot spin-qubit in a waveguide-based spin-photon interface, Phys. Rev. Applied **11**, 031002 (2019).

[15] Ł. Dusanowski, C. Nawrath, S. L. Portalupi, M. Jetter, T. Huber, S. Klembt, P. Michler, and S. Höfling, Optical charge injection and coherent control of a quantum-dot spin-qubit emitting at telecom wavelengths, Nat. Commun. **13**, 748 (2022).

[16] D. Press, K. De Greve, P. L. McMahon, T. D. Ladd, B. Friess, C. Schneider, M. Kamp, S. Höfling, A. Forchel, and Y. Yamamoto, Ultrafast optical spin echo in a single quantum dot, Nat. Photon. **4**, 367 (2010).

[17] S. T. Yılmaz, P. Fallahi, and A. Imamoğlu, Quantum-dot-spin single-photon interface, Phys. Rev. Lett. **105**, 033601 (2010).

[18] M. H. Appel, A. Tiranov, A. Javadi, M. C. Löbl, Y. Wang, S. Scholz, A. D. Wieck, A. Ludwig, R. J. Warburton, and P. Lodahl, Coherent spin-photon interface with waveguide induced cycling transitions, Phys. Rev. Lett. **126**, 013602 (2021).

[19] K. De Greve, L. Yu, P. L. McMahon, J. S. Pelc, C. M. Natarajan, N. Y. Kim, E. Abe, S. Maier, C. Schneider, M. Kamp, S. Höfling, R. H. Hadfield, A. Forchel, M. M. Fejer, and Y. Yamamoto, Quantum-dot spin–photon entanglement via frequency downconversion to telecom wavelength, Nat. **491**, 421 (2012).

[20] W. B. Gao, P. Fallahi, E. Togan, J. Miguel-Sanchez, and A. Imamoglu, Observation of entanglement between a quantum dot spin and a single photon, Nat. **491**, 426 (2012).

[21] M. Atatüre, J. Dreiser, A. Badolato, and A. Imamoglu, Observation of Faraday rotation from a single confined spin, Nat. Phys. **3**, 101 (2007).

[22] A. J. Ramsay, T. M. Godden, S. J. Boyle, E. M. Gauger, A. Nazir, B. W. Lovett, A. V. Gopal, A. M. Fox, and M. S. Skolnick, Effect of detuning on the phonon induced dephasing of optically driven InGaAs/GaAs quantum dots, J. Appl. Phys. **109**, 102415 (2011).

[23] S. A. Crooker, J. Brandt, C. Sandfort, A. Greilich, D. R. Yakovlev, D. Reuter, A. D. Wieck, and M. Bayer, Spin noise of electrons and holes in self-assembled quantum dots, Phys. Rev. Lett. **104**, 036601 (2010).

[24] K. Barr, B. Hourahine, C. Schneider, S. Höfling, and K. G. Lagoudakis, Towards spin state tailoring of charged excitons in InGaAs quantum dots using oblique magnetic fields, Phys. Rev. B **109**, 075433 (2024).

[25] E. Poem, J. Shemesh, I. Marderfeld, D. Galushko, N. Akopian, D. Gershoni, B. D. Gerardot, A. Badolato, and P. M. Petroff, Polarization sensitive spectroscopy of charged quantum dots, Phys. Rev. B **76**, 235304 (2007).

[26] A. Schwan, B.-M. Meiners, A. Greilich, D. R. Yakovlev, M. Bayer, A. D. B. Maia, A. A. Quivy, and A. B. Henriques, Anisotropy of electron and hole $g$-factors in (In,Ga)As quantum dots, Appl. Phys. Lett. **99**, 221914 (2011).

[27] I. Hapke-Wurst, U. Zeitler, R. Haug, and K. Pierz, Mapping the g factor anisotropy of InAs self-assembled quantum dots, Physica E: Low Dimens. Syst. Nanostruct. **12**, 802 (2002).

[28] V. V. Belykh, A. Greilich, D. R. Yakovlev, M. Yacob, J. P. Reithmaier, M. Benyoucef, and M. Bayer, Electron and hole $g$ factors in InAs/InAlGaAs self-assembled quantum dots emitting at telecom wavelengths, Phys. Rev. B **92**, 165307 (2015).

[29] B. Van Hattem, P. Corfdir, P. Brereton, P. Pearce, A. Graham, M. Stanley, M. Hugues, M. Hopkinson, and R. Phillips, Photoluminescence in tilted magnetic field of triply negatively charged excitons hybridized with a continuum, Acta Phys. Pol. A **124**, 798 (2013).

[30] M. Neumann, F. Kappe, T. K. Bracht, M. Cosacchi, T. Seidelmann, V. M. Axt, G. Weihs, and D. E. Reiter, Optical stark shift to control the dark exciton occupation of a quantum dot in a tilted magnetic field, Phys. Rev. B **104**, 075428 (2021).

[31] C. A. Jiménez-Orjuela, H. Vinck-Posada, and J. M. Villas-Bôas, Dark excitons in a quantum-dot–cavity system under a tilted magnetic field, Phys. Rev. B **96**, 125303 (2017).

[32] S. Maier, P. Gold, A. Forchel, N. Gregersen, J. Mørk, S. Höfling, C. Schneider, and M. Kamp, Bright single photon source based on self-aligned quantum dot–cavity systems, Opt. Express **22**, 8136 (2014).

[33] K. Barr, T. Cookson, and K. G. Lagoudakis, Operation of a continuous flow liquid helium magnetic microscopy

cryostat as a closed cycle system, Rev. Sci. Instrum. **92**, 123701 (2021).
[34] T. A. Wilkinson, D. J. Cottrill, J. M. Cramlet, C. E. Maurer, C. J. Flood, A. S. Bracker, M. Yakes, D. Gammon, and E. B. Flagg, Spin-selective AC Stark shifts in a charged quantum dot, Appl. Phys. Lett. **114**, 133104 (2019).
[35] J. Johansson, P. Nation, and F. Nori, QuTiP: An open-source Python framework for the dynamics of open quantum systems, Comput. Phys. Commun. **183**, 1760 (2012).
[36] J. Johansson, P. Nation, and F. Nori, QuTiP 2: A Python framework for the dynamics of open quantum systems, Comput. Phys. Commun. **184**, 1234 (2013).
[37] T. Okada, K. Komori, K. Goshima, S. Yamauchi, I. Morohashi, T. Sugaya, M. Ogura, and N. Tsurumachi, Development of high resolution Michelson interferometer for stable phase-locked ultrashort pulse pair generation, Rev. Sci. Instrum. **79**, 103101 (2008).
[38] M. Kotur, P. S. Bazhin, K. V. Kavokin, N. E. Kopteva, D. R. Yakovlev, D. Kudlacik, and M. Bayer, Dynamic polarization of nuclear spins by optically oriented electrons and holes in lead halide perovskite semiconductors, Phys. Rev. B **113**, 085204 (2026).
[39] S. E. Economou and E. Barnes, Theory of dynamic nuclear polarization and feedback in quantum dots, Phys. Rev. B **89**, 165301 (2014).
[40] S. Yamamoto, R. Kaji, H. Sasakura, and S. Adachi, Effect of oblique magnetic field on dynamic nuclear polarization in single self-assembled quantum dots governed by nuclear quadrupole interaction, Phys. stat. sol. (b) **262**, 2400535 (2025).
[41] T. D. Ladd, D. Press, K. De Greve, P. L. McMahon, B. Friess, C. Schneider, M. Kamp, S. Höfling, A. Forchel, and Y. Yamamoto, Pulsed nuclear pumping and spin diffusion in a single charged quantum dot, Phys. Rev. Lett. **105**, 107401 (2010).
[42] P. Millington-Hotze, H. E. Dyte, S. Manna, S. F. Covre da Silva, A. Rastelli, and E. A. Chekhovich, Approaching a fully-polarized state of nuclear spins in a solid, Nat. Commun. **15**, 985 (2024).
[43] Numerical parameters used in the simulations for the Oblique (Voigt) configuration, $\frac{\Delta E_{gs}}{2\pi} = 32 GHz (31 GHz)$, $\frac{\Delta E_{es}}{2\pi} = 63.4 GHz (9.7 GHz)$, $\frac{\sigma}{2\pi} = 98 GHz$ which correspond to a width of $\approx$ 3psec pulse, $\frac{\Omega_{Rabi}^{CW}}{2\pi} = 1 GHz$, $\frac{\Omega_{Ramsey}^{CW}}{2\pi} = \frac{\Omega_{Rabi}^{CW}}{\sqrt{3}}$, $\frac{\Omega_{SU(2)}^{CW}}{2\pi} = \frac{\Omega_{Rabi}^{CW}}{2}$, $\frac{\Omega^{Pulsed}}{2\pi} \in (0, 2550) GHz$, $\frac{\Delta_{Pulsed}}{2\pi} = -500 GHz$, $\Gamma = k_{ij}^2$, $\gamma_{dephasing} = 1 ns^{-1}$, $\alpha_{phonon} = 28 * 10^{-6} ns^{-1}$, $\vartheta = 60°(90°)$, $\alpha = 90°(45°)$, $\beta = 10°(90°)$, $k_{14} = k_{23} = 0.75(0.5)$, $k_{24} = k_{13} = 0.25(0.5)$.

# Complete coherent control of spin qubits in self-assembled InAs quantum dots under oblique magnetic fields: supplement

## A. Abstract

This document is the supplementary material of the manuscript entitled "Complete coherent control of spin qubits in self-assembled InAs quantum dots under oblique magnetic fields", where we present additional measurements and analysis for a self-assembled InAs quantum dot under magnetic field. Specifically, we report data on the Zeeman splitting, the diamagnetic shift, and the optical dressing of the transition, which together further characterise the energy structure and driven response of the system. In addition, a significant part of this Supplemental Material is dedicated to nuclear polarization, its experimental manifestations, and its influence on the observed dynamics. The material presented here provides further detail and support for the results and interpretation discussed in the main text.

## B. Zeeman interaction & dressing of the driven transition

### Zeeman splitting

Figure S1(a) shows the evolution of the optical transition energies between the ground-spin state manifold and the trion manifold as a function of magnetic field. This dependence arises from the interplay between the Zeeman interaction, which shifts the energies linearly with magnetic field and lifts the spin-state degeneracy, and the diamagnetic shift, which contributes a quadratic-in-field correction to the transition energies. As a result, the measured spectral lines exhibit a characteristic field dependence that cannot be described by a purely linear model alone. By simultaneously fitting the magnetic-field dependence of the four optical transitions, we extract the diamagnetic contribution in a consistent manner and obtain a diamagnetic coefficient of $\gamma = 6.021\,\mu\mathrm{eV}/\mathrm{T}^2$, as presented in Fig. S1(b). Figures S1(c), (d) show the Voigt equivalent of Figures S1(a), (b) correspondingly, where we report a a diamagnetic coefficient of $\gamma = 4.395\,\mu\mathrm{eV}/\mathrm{T}^2$ in the Voigt case.

### Dressing of the resonantly driven transition

Figure S2(a) presents the driven transition as a function of the applied CW excitation power. With increasing CW power, clear signatures of optical dressing emerge only for powers above $\sim 2.5\,\mu\mathrm{W}$, indicating the onset of a substantial modification of the transition S2(b) by the driving field. Because the power range used in the coherent-control experiments lies well below this threshold, the observed coherent control cannot be attributed to CW dressing effects and instead reflects the system's intrinsic low-power quantum coherent response.

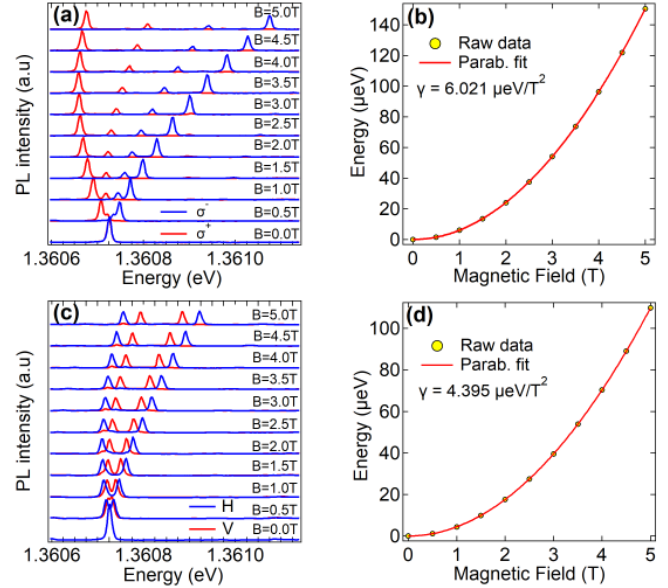

FIG. S1. (a) Raw photoluminescence spectra recorded at different magnetic-fields strengths B in the Oblique geometry. (b) Quadratic field dependence of the transition energies attributed to the diamagnetic shift (fit). (c) PL spectra at different B in Voigt. (d) Diamagnetic shift fit in the Voigt geometry.

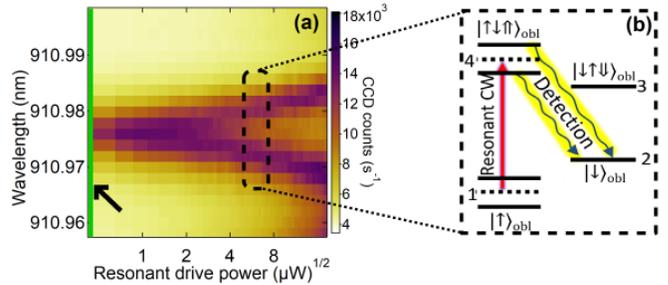

FIG. S2. (a) Raw PL spectra recorded for increasing CW power, with the x-axis displayed in logarithmic scale. At $\sim 2.5\,\mu\mathrm{W}$ we observe dressed states caused by the CW power. The black arrow indicates the green shaded area where we operate in the coherent control experiment. (b) Double $\Lambda$-system at $\sim 6.5\,\mu\mathrm{W}$ in order for the reader to understand how this phenomenon affect our system.

## C. Oblique magnetic field's nuclear polarization

During the acquisition of Rabi-oscillation and Ramsey-interference measurements, we observed that the coherence properties depend reproducibly on the detuning of the resonant continuous-wave (CW) laser from the targeted optical transition. In particular, the form of the Rabi oscillations changes systematically with CW detuning: by adjusting the detuning, the signal can be made to evolve slightly along either an upward or a downward trend (Fig. S3(a)). This behavior was observed consistently while stepping the CW laser frequency across the transition and repeating the pulsed-control measurements (Fig. S3(b)), suggesting that the nuclear environment is not merely a passive source of quasi-static noise, but is instead influenced by the optical driving conditions. In addition, when the CW laser is set slightly off resonance, while keeping the pulse parameters and detection configuration unchanged, the Ramsey fringe visibility can increase relative to exact resonance (Fig. S4). We attribute this detuning dependence to detuning-controlled dynamic nuclear polarization (DNP) generated by resonant optical cycling. In a charged quantum dot, repeated excitation and spontaneous emission can produce a non-equilibrium carrier-spin polarization, which couples to the host nuclei via the hyperfine interaction. As a result, the CW laser may drive the nuclear bath into different polarization states depending on its detuning and scan history, thereby modifying the effective Overhauser field experienced by the confined carrier [1–3]. Within this picture, changing the CW detuning impacts the induced nuclear polarization, which can in turn shift the effective transition frequency through an Overhauser shift and modify the dephasing landscape relevant to the Ramsey measurements. We emphasize that, in the present dataset, this conclusion is based on a consistent experimental correlation rather than on a direct measurement of the Overhauser-field distribution. Alternative, non-nuclear contributions, such as detuning-dependent power broadening, residual repumping efficiency, AC-Stark-like shifts arising from imperfect spectral filtering, or slow charge-noise drift correlated with laser frequency, cannot be fully excluded. Nevertheless, the observed sensitivity of both the Rabi response and the Ramsey coherence to small changes in CW detuning, obtained while repeating identical pulsed-control sequences as the CW frequency is set to different values, strongly supports the working hypothesis that CW detuning has a direct impact on the nuclear spin bath polarization. This also suggests that an appropriate choice of detuning may provide a practical route to maximizing coherence by mitigating nuclear-induced dephasing.

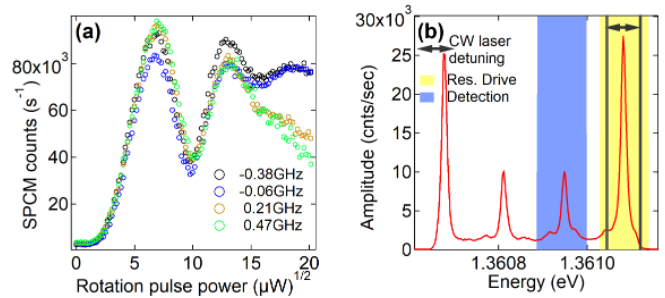

FIG. S3. (a) Rabi Oscillations for different detuning of the CW laser as a function of the pulse power. (b) Quantum dot spectrum showing the frequency scan of the CW laser across the targeted optical transition, with the black vertical lines indicating the detuning range used in the experiment.

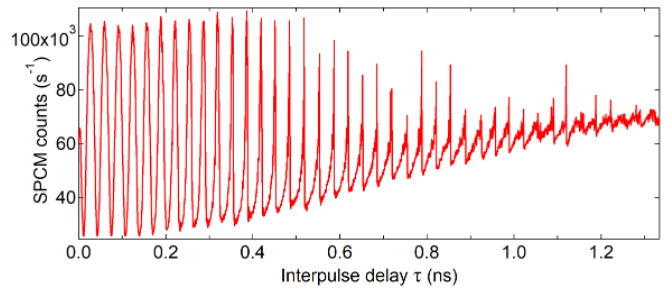

FIG. S4. Ramsey fringes for a slightly modified resonant-laser detuning ($\sim -0.4\,\text{GHz}$ from "resonance"), showing enhanced coherence times.

## REFERENCES

[1] T. D. Ladd, D. Press, K. De Greve, P. L. McMahon, B. Friess, C. Schneider, M. Kamp, S. Höfling, A. Forchel, and Y. Yamamoto, Pulsed nuclear pumping and spin diffusion in a single charged quantum dot, Phys. Rev. Lett. 105, 107401 (2010).

[2] S. E. Economou and E. Barnes, Theory of dynamic nuclear polarization and feedback in quantum dots, Phys. Rev. B 89, 165301 (2014).

[3] P. Millington-Hotze, H. E. Dyte, S. Manna, S. F. Covre da Silva, A. Rastelli, and E. A. Chekhovich, Approaching a fully-polarized state of nuclear spins in a solid, Nat Commun 15, 985 (2024).